\documentclass[letterpaper]{article} 
\usepackage{aaai2026}  
\usepackage{times}  
\usepackage{helvet}  
\usepackage{courier}  
\usepackage[hyphens]{url}  
\usepackage{graphicx} 
\usepackage{natbib}  
\usepackage{caption} 
\usepackage{algorithm}
\usepackage{algorithmic}

\usepackage{newfloat}
\usepackage{listings}
\DeclareCaptionStyle{ruled}{labelfont=normalfont,labelsep=colon,strut=off} 
\floatstyle{ruled}
\newfloat{listing}{tb}{lst}{}
\floatname{listing}{Listing}
\title{Multimodal Large Language Models as Synthetic Participants\\ in Video-Based Studies: An Evaluation}
\author{
    Prabal Shrestha\textsuperscript{\rm 1}, Bohan Jiang\textsuperscript{\rm 2}, Haoning Xue\textsuperscript{\rm 3}, Huan Liu\textsuperscript{\rm 2}, Xinyi Zhou\textsuperscript{\rm 1}\thanks{Corresponding author.}}
\affiliations{
    \textsuperscript{\rm 1}Department of Computer Science, Boise State University, Boise, ID, USA \\
    \textsuperscript{\rm 2}School of Computing and Augmented Intelligence, Arizona State University, Tempe, AZ, USA \\
    \textsuperscript{\rm 3}Department of Communication, University of Utah, Salt Lake City, UT, USA \\

    prabalshrestha@u.boisestate.edu,
    bjiang14@asu.edu,
    haoning.xue@utah.edu,
    huanliu@asu.edu,
    xinyizhou@boisestate.edu
}

\usepackage{bibentry}

\usepackage{amssymb}
\usepackage{booktabs}
\usepackage{multirow}
\usepackage{subcaption}
\usepackage{makecell}
\usepackage[most]{tcolorbox}
\usepackage{xcolor}
\usepackage{adjustbox}

\begin{document}

\maketitle

\begin{figure*}[t]
\centering
\includegraphics[width=\textwidth]{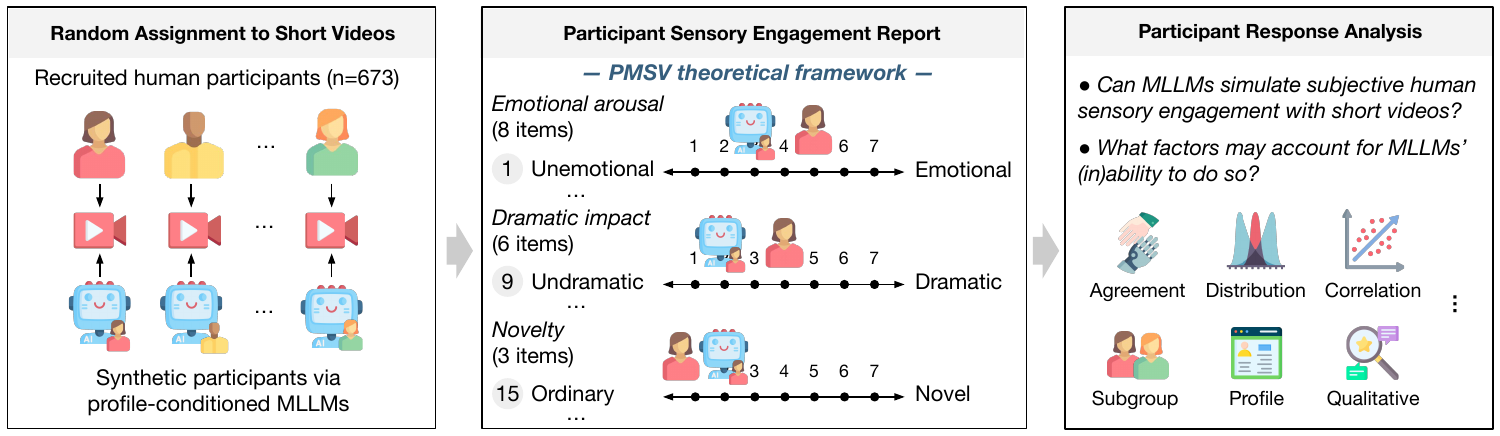}
\caption{Study overview.}
\label{fig:overview}
\end{figure*}

\begin{abstract}
Multimodal large language models (MLLMs) have shown strong performance on objective tasks such as video understanding and reasoning. However, it remains unclear whether they can approximate subjective human responses, which depend not only on content comprehension but also on individuals' social contexts. To address this gap, we evaluate MLLMs as synthetic participants in an emerging task: assessing perceived sensory engagement with short videos.
Grounded in the Perceived Message Sensation Value (PMSV) framework, we compare ratings from recruited human participants and profile-conditioned MLLM simulations ($n=673$) using a 17-item scale measuring emotional arousal, dramatic impact, and novelty. We find that even leading MLLMs (Gemini 3 Flash and Qwen 3 Omni) show limited agreement with human participants. The models exhibit distinct downward mean-shift and central-tendency biases in their rating distributions. They both introduce and flatten subgroup differences, while showing inconsistent sensitivity to participant profiles. Prompting strategies affect these metrics differently, modestly improving some aspects while worsening others. These results highlight both the challenges and opportunities of developing MLLMs as synthetic participants in video-based research.
\end{abstract}

\begin{links}
    \link{Data and code}{https://github.com/MINDLab25/mllm-human-simulation-eval}
\end{links}

\section{Introduction}
Human-subject studies are essential to science, policy, and medicine, but they are time-consuming, labor-intensive, and need careful risk management, especially when they involve vulnerable groups~\cite{uybico2007recruiting,o2022rigor}. Large language models (LLMs) offer a potential way to scale and strengthen human-subject studies by simulating a range of participants and their likely responses, helping researchers refine protocols, identify potential failure points, and surface possible risks earlier in the process.

In parallel, multimodal LLMs (MLLMs) have demonstrated strong capabilities in \textit{objective} understanding and reasoning over multimodal inputs, including video content~\cite{yue2024mmmu,xu2025qwen3}, raising the question of whether they can also simulate how people respond to such content. However, human responses are inherently \textit{subjective}, shaped not only by content comprehension but also by background, interpretation, preferences, and other contextual factors. Strong performance on objective tasks therefore does not necessarily imply an ability to approximate subjective human judgments, and whether MLLMs can bridge this gap, particularly for video stimuli, remains unexplored.

Here, we investigate:
(\textbf{RQ1}) \textit{To what extent can advanced MLLMs simulate subjective human responses in video-based research?} 
and
(\textbf{RQ2}) \textit{What factors may account for their (in)ability to do so?}
We focus on human sensory engagement with short video content, a domain of growing societal relevance (Figure~\ref{fig:overview}). Short video platforms have been major channels for information consumption, and their recommendation systems prioritize content maximizing behavioral engagement, which is closely tied to viewers' sensory engagement with the content~\cite{zannettou2024analyzing}. Understanding how individuals perceive and respond to short videos is therefore critical, as these signals shape recommendation dynamics, content diffusion, and broader societal impact. 

Drawing on PMSV, a theoretical framework~\cite{palmgreen2002perceived}, we operationalize viewers' sensory engagement with short-form video content using a 17-item scale spanning emotional arousal, dramatic impact, and novelty. 673 human participants reported their perceptions on all items after watching randomly assigned videos from \citet{xue2026msv}, along with their demographic and psychographic profiles. Conditioned on these profiles, MLLMs (Gemini 3 Flash and Qwen 3 Omni) predict human responses using prompting-based methods.

The results suggest that although leading MLLMs can often understand and reason about video content, they struggle to simulate individuals' subjective sensory engagement with such content, showing limited agreement with human participants that is substantially below a human--human agreement reference. 
The models exhibit distinct downward mean-shift and central-tendency biases in their rating distributions.
They also distort subgroup-level patterns, at times introducing artificial differences and at other times flattening differences observed in the human data, while showing inconsistent sensitivity to participant profiles. 
Prompting strategies affect these metrics differently, modestly improving some aspects while worsening others. 
This work provides empirical evidence and diagnostic insights for developing MLLMs as synthetic participants in video-based studies.

\section{Related Work}

Recent work has explored LLMs as synthetic participants for surveys, experiments, and social-behavioral simulation. Early studies argued that, when conditioned on demographic information, LLMs can achieve algorithmic fidelity~\cite{ma2025algorithmic} by reproducing some response patterns of human subpopulations in surveys~\cite{argyle2023out}. More recent work has shown that task-specific adaptation improves simulation quality beyond zero-shot prompting~\cite{kolluri2025finetuning}. However, most prior work has focused on text-based social experiments and simulations~\cite{manning2024automated}. Our paper extends this line of research to a more challenging and important setting: simulating human responses to video, where reactions depend jointly on visual, auditory, and textual cues.

A parallel literature has examined whether MLLMs can understand and reason about multimodal content. Recent Gemini and Qwen models have shown strong performance on video understanding and reasoning~\cite{xu2025qwen3}. Other work has introduced benchmarks for recognizing human values in video data, arguing that socially grounded video understanding is an important but underexplored capability for MLLMs~\cite{wang2025multimodal}. In addition, prior studies suggest that MLLMs can be aligned with human social judgments through training on behavioral annotations~\cite{garcia2025aligning,fleury2024video}. Our work builds on this literature but shifts the focus from general multimodal understanding and reasoning to subjective human response simulation, using PMSV as a psychometrically motivated outcome in short video contexts.

\section{Experimental Setup \& Results}

We randomly sampled 120 short-form videos from \citet{xue2026msv}. The videos spanned diverse topics, were published between August 2020 and April 2022 by 67 unique Instagram accounts with varying levels of popularity, and had an average duration of 44.8 seconds.  
We recruited US adults via Prolific; each participant watched a randomly assigned video and reported their sensory engagement using a 17-item PMSV scale (1--7) covering emotional arousal, dramatic impact, and novelty (Appendix). Participants also reported demographic information and completed the brief sensation-seeking scale~\cite{hoyle2002reliability} (Appendix), because sensation seeking has been shown to influence sensory engagement with media content~\cite{palmgreen2002perceived}. 

A total of 673 participants passed the attention checks and were included in the final analysis. Each participant evaluated two videos, and each video received eight \textit{independent} evaluations, on average. We computed rating-level PMSV as the mean of the 17 PMSV items~\cite{palmgreen2002perceived}, participant-level PMSV by averaging the rating-level PMSV scores across the videos evaluated by each participant. 

We evaluated two MLLMs: Gemini 3 Flash (commercial, hereafter, ``Gemini'') and Qwen 3 Omni (open-source, hereafter, ``Qwen''). 
Both support raw video input, including textual, visual, and audio modalities, and were state-of-the-art video-capable MLLMs as of January 2026. 
For each participant--video pair, we prompted each MLLM with the assigned video and the participant profile, including all reported demographics and the sensation-seeking score, and asked it to predict the participant's ratings on the 17 PMSV items (Appendix). 
We then computed synthesized rating-level PMSV from the predicted item ratings and averaged across the assigned videos to obtain synthesized participant-level PMSV, mirroring the human PMSV construction.

We considered three prompting strategies: zero-shot, few-shot, and CoT prompting (Appendix, adapted from \citet{kolluri2025finetuning}). 
For the few-shot prompt, we followed \citet{min2022rethinking}'s guidance and included three examples: three videos that differed in topic and length and were excluded from the evaluation set, each paired with one human participant's survey response. 
Together, the three example participants spanned the profile characteristics used in the MLLM simulation. 
See the Appendix for implementation details.

\begin{figure}[t]
    \centering
    \includegraphics[width=\linewidth]{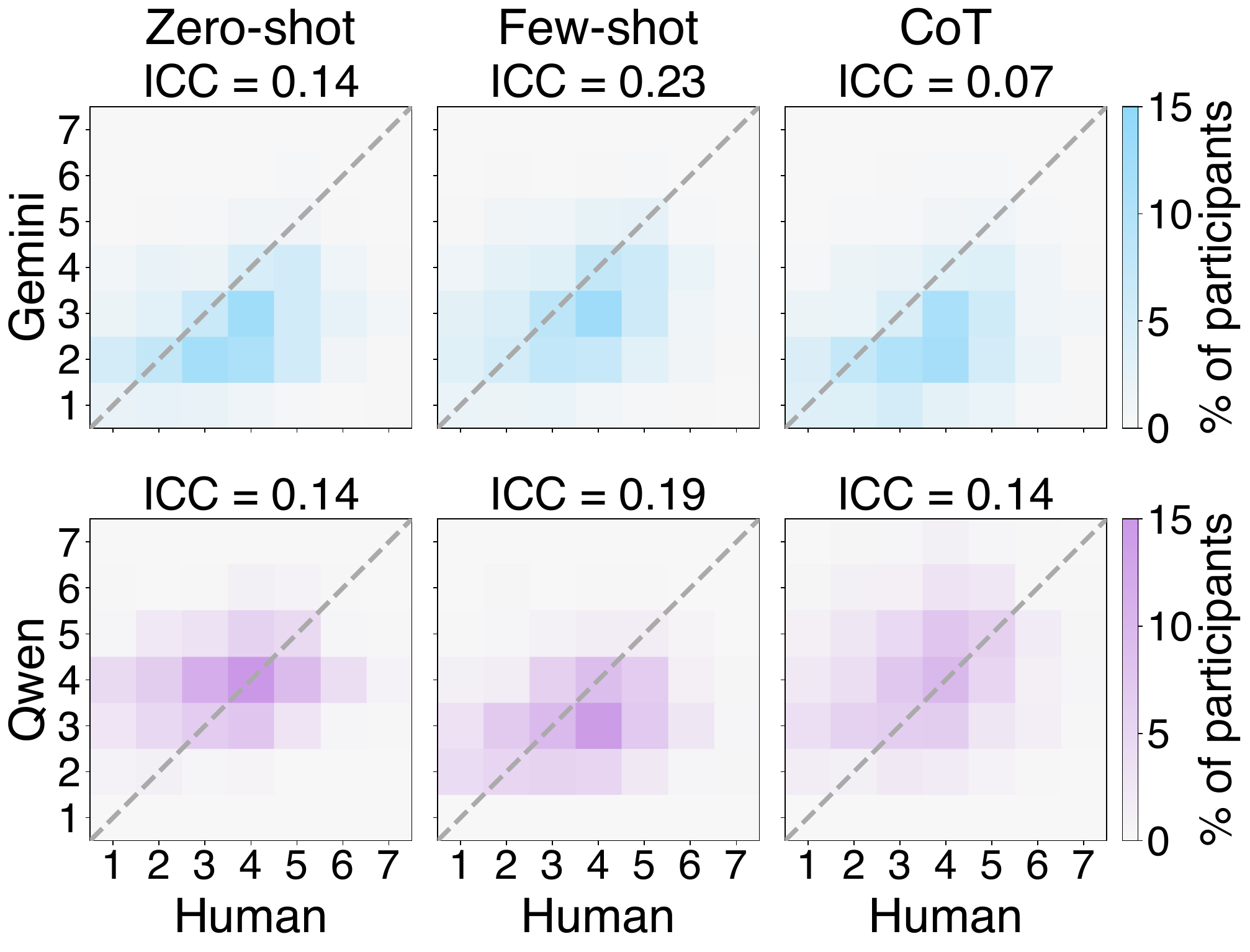}
    \caption{PMSV agreement between human and matched MLLM-synthesized participants across prompting methods.}
    \label{fig:pmsv_agmt}
\end{figure}

\begin{table}[t]
\adjustbox{max width=\textwidth}{
\centering \small
\begin{tabular}{rlcc}
&  & \textbf{Gemini} & \textbf{Qwen} \\ \midrule
\textbf{Gender} &
  Male (318) & 0.22 & 0.16 \\
& Female (337) & 0.07 & 0.15 \\
\textbf{Ethnicity} &
  White (466) & 0.13 & 0.14 \\
& Black (88) & 0.08 & 0.20 \\
& Asian (43) & 0.17 & 0.24 \\
\textbf{Age} & 
$<39$ (338) & 0.24 & 0.16 \\
& $\geq 39$ (334) & 0.05 & 0.13 \\
\textbf{Education} & 
$<$ Bachelor's (324) & 0.17 & 0.13 \\
& $\geq$ Bachelor's (346) & 0.12 & 0.16 \\
\textbf{Household income} & 
$<\$$50k (253) & 0.15 & 0.11 \\
& $\$$50k--$\$$100k (241) & 0.20 & 0.23 \\
& $\geq\$$100k (156) & 0.04 & 0.05 \\
\textbf{Sensation seeking} & 
$<2.5$ (312) & 0.02 & 0.14 \\
& $\geq 2.5$ (360) & 0.22 & 0.13 \\
\end{tabular}}
\caption{PMSV agreement between human and matched MLLM-synthesized participants within each subgroup (zero-shot). Median splits: age (39), sensation seeking (2.5).}
\label{tab:pmsv_subgroups}
\end{table}

\smallskip
\begin{tcolorbox}[
  title={$\blacktriangleright$ Finding 1},
  detach title,
  coltitle=black,
  fonttitle=\bfseries,
  before upper={%
    \textbf{\tcbtitletext}:\enskip
  },
  top=2pt, bottom=2pt,
  left=2pt, right=2pt,
  arc=0pt, outer arc=0pt,
  boxrule=1pt
]
  MLLMs exhibit limited individual-level agreement with human participants.
\end{tcolorbox}

Under zero-shot prompting, agreement between each human participant's PMSV and that of the corresponding MLLM-synthesized participant---measured using the single-measure, absolute-agreement intraclass correlation coefficient (hereafter, ``ICC'')---is 0.14 for both Gemini and Qwen (Figure~\ref{fig:pmsv_agmt}). 
This MLLM--human agreement is substantially below the human--human agreement reference of 0.29 (95\% CI = [0.269, 0.305]; Appendix), indicating that current profile-conditioned MLLMs remain below the level of individual agreement observed among humans when simulating sensory engagement with short video content. 
This low agreement persists across demographic and psychographic subgroups for both MLLMs: ICC ranges from 0.02 to 0.24 for Gemini and from 0.05 to 0.24 for Qwen (Table~\ref{tab:pmsv_subgroups}). 

Few-shot prompting modestly increases MLLM--human agreement, with ICC rising to 0.23 for Gemini ($\Delta_{\mathrm{ICC}}$~=~0.09) and to 0.19 for Qwen ($\Delta_{\mathrm{ICC}}$~=~0.05), although both remain below the human--human reference. By contrast, CoT prompting does not improve agreement: ICC decreases to 0.07 for Gemini ($\Delta_{\mathrm{ICC}}$~=~-0.07) and remains unchanged at 0.14 for Qwen (Figure~\ref{fig:pmsv_agmt}), despite being more resource-intensive. The same pattern is reflected in MAE computed over matched human--MLLM pairs (Appendix). 
One possible explanation is that CoT prompting is misaligned with the PMSV task: it encourages explicit step-by-step deliberation~\cite{wei2022chain}, whereas PMSV judgments may reflect fast, sensory, and subjective impressions. CoT may therefore shift the model away from a simulated subjective ``participant'' and toward a more analytical evaluator.

\begin{figure}[t]
\centering
    \includegraphics[width=\linewidth]{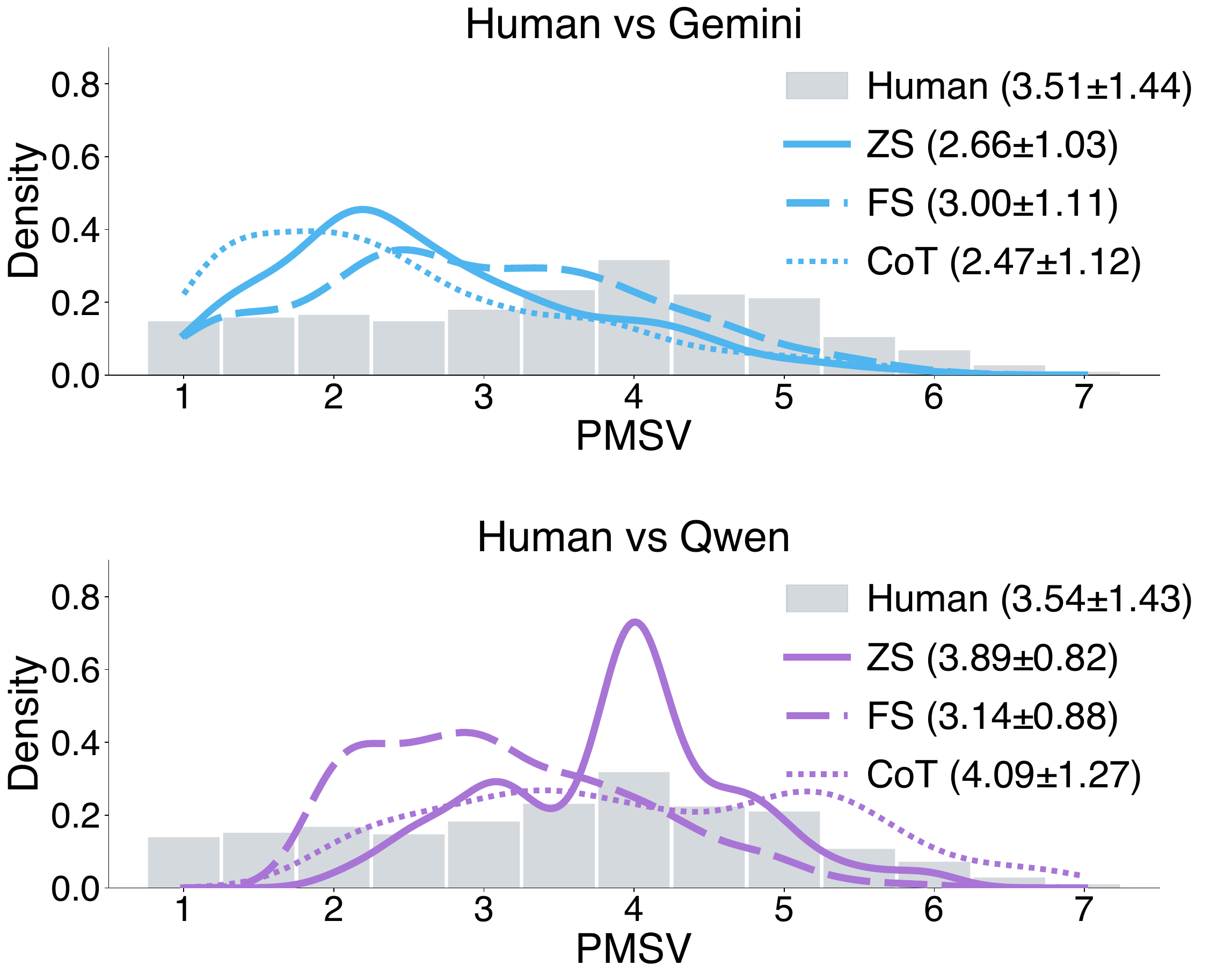}
    \caption{Distributions of rating-level human and MLLM-simulated PMSVs across prompting methods. ZS: zero-shot; FS: few-shot.}
    \label{fig:pmsv_hist}
\end{figure}

\smallskip
\begin{tcolorbox}[
  title={$\blacktriangleright$ Finding 2},
  detach title,
  coltitle=black,
  fonttitle=\bfseries,
  before upper={%
    \textbf{\tcbtitletext}:\enskip
  },
  top=2pt, bottom=2pt,
  left=2pt, right=2pt,
  arc=0pt, outer arc=0pt,
  boxrule=1pt
]
  MLLMs show distinct biases in rating distributions: downward mean-shift vs central tendency.
\end{tcolorbox}

Under zero-shot prompting, both Gemini and Qwen show systematic biases in predicting rating-level human PMSV, but the nature of these biases differs (Figure~\ref{fig:pmsv_hist}). 
Qwen primarily shows a \textit{central tendency bias}, with predictions clustered near the scale midpoint. 
As a result, its outputs are less variable than human ratings (Qwen SD~=~0.82 vs human SD~=~1.43; $p<.001$; LMM on squared deviations with source as a fixed effect and participant as a random intercept).
This pattern is consistent with findings from \citet{hong2026rulers}, an LLM-as-a-judge study on text-based tasks. 

Gemini primarily presents a \textit{downward mean-shift bias}: its predictions are shifted toward lower scores, with a mean rating-level PMSV of 2.66, significantly below the human mean of 3.51 ($p<.001$; LMM with source as a fixed effect and participant as a random intercept; Appendix). This downward shift persists even when the rating order of the PMSV items is reversed (e.g., from emotional--unemotional to unemotional--emotional), suggesting that it is not simply an artifact of item-anchor direction.

Distributional bias and relative ordering capture different aspects of simulation performance. 
Despite its downward mean shift, Gemini better preserves the relative ordering of human PMSV scores than Qwen: its Pearson correlation with human rating-level PMSV is 0.38, compared with Qwen's 0.24. 
Thus, Gemini can partially track which participant--video responses are relatively higher or lower, while still failing to match humans in absolute agreement, as shown in Finding~1.

Prompting strategies affect these biases differently. 
Figure~\ref{fig:pmsv_hist} shows that few-shot and CoT prompting mitigate, but do not eliminate, central tendency bias. 
However, downward mean-shift bias remains difficult to mitigate, and CoT prompting exacerbates it in both MLLMs. 
One possible explanation is that internal reasoning may fail to correct initial biases and can reinforce them in the absence of external feedback~\cite{huang2023large}.

\begin{figure}[t]
    \centering
    \begin{subfigure}{\linewidth}
        \includegraphics[width=\linewidth]{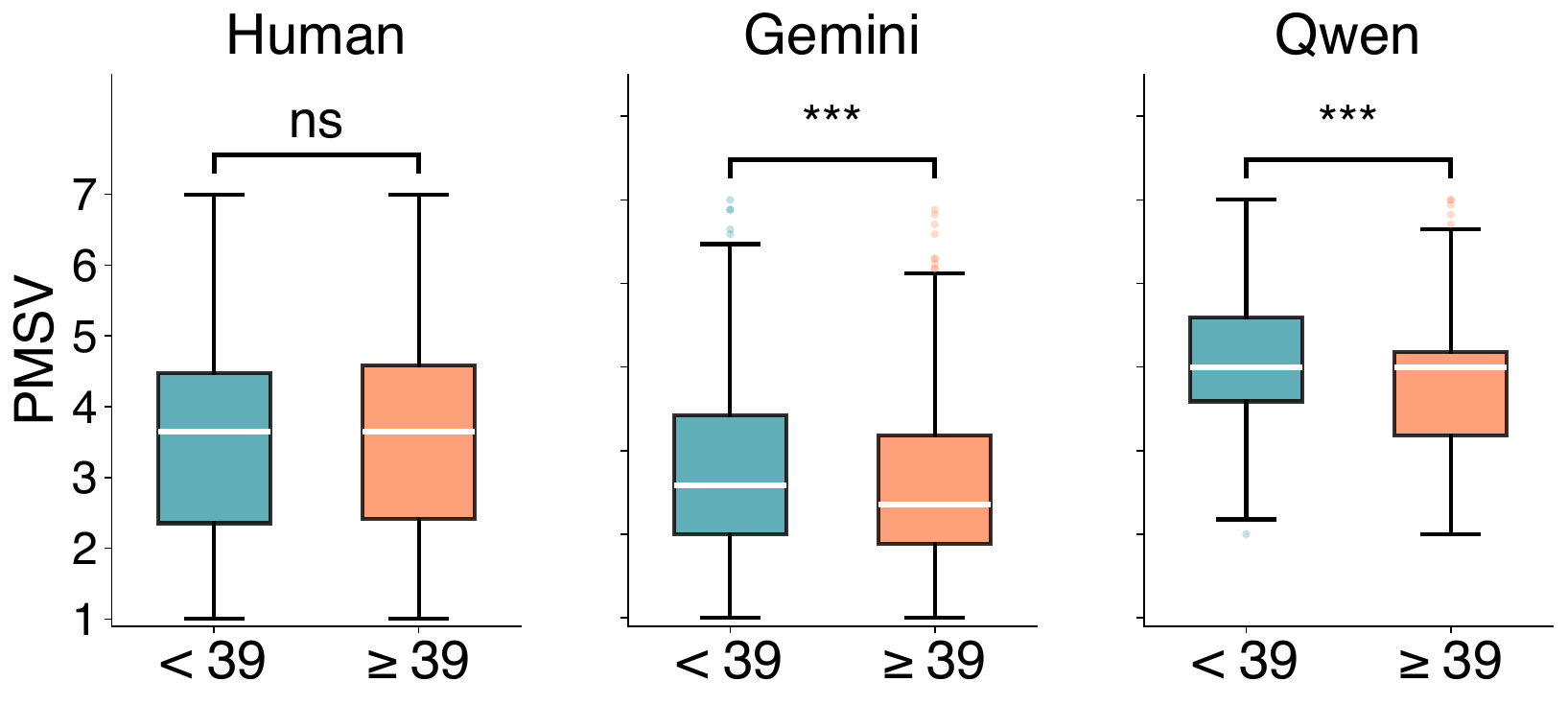}
        \caption{Age}
        \label{fig:pmsv_age}
    \end{subfigure}\hfill
    \begin{subfigure}{\linewidth}
        \includegraphics[width=\linewidth]{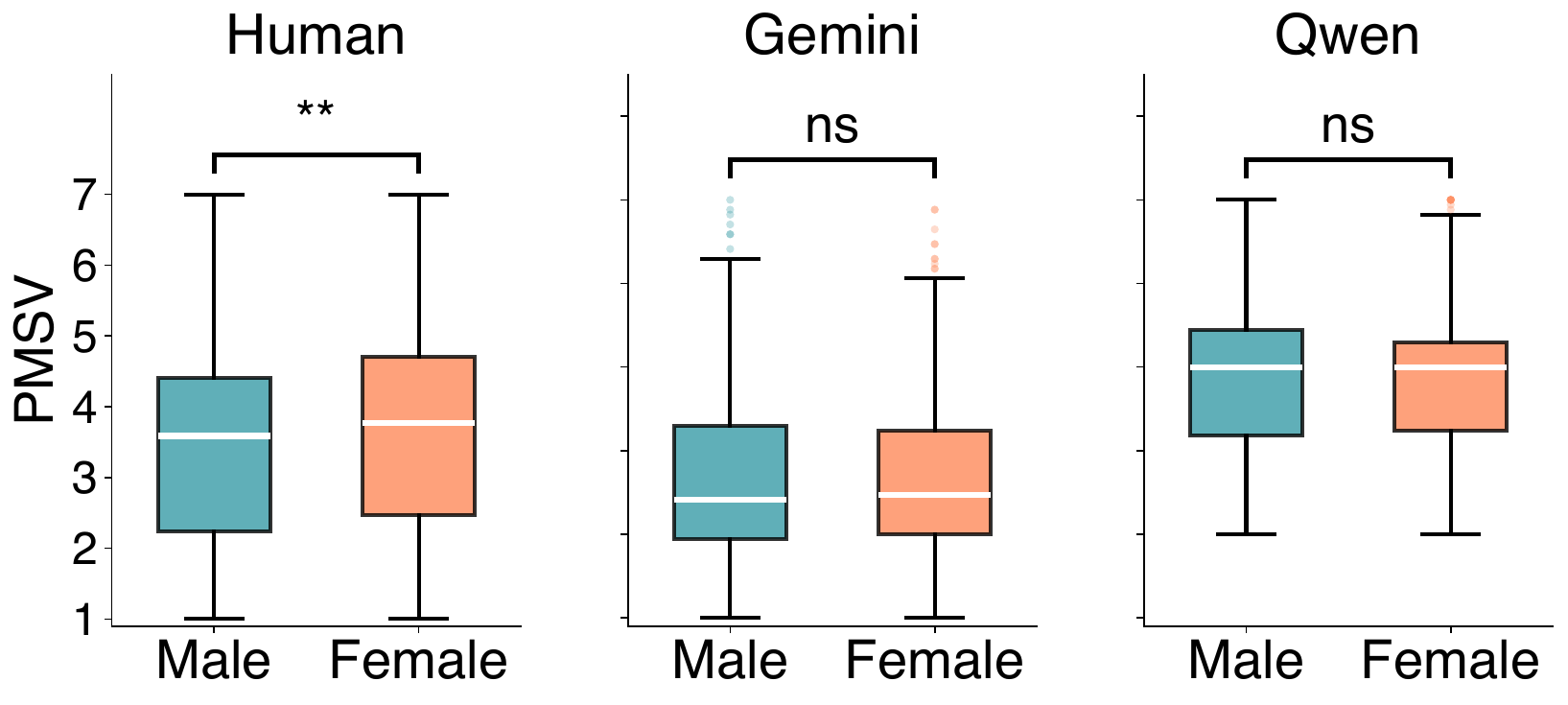}
        \caption{Gender}
        \label{fig:pmsv_gender}
    \end{subfigure}
    \caption{PMSV distributions by group (zero-shot). $^{***}$: $p<.001$, $^{**}$: $p<.01$, ns: $p\geq .05$; $t$-test. Age: median split at 39 years.}
\end{figure}

\smallskip
\begin{tcolorbox}[
  title={$\blacktriangleright$ Finding 3},
  detach title,
  coltitle=black,
  fonttitle=\bfseries,
  before upper={%
    \textbf{\tcbtitletext}:\enskip
  },
  top=2pt, bottom=2pt,
  left=2pt, right=2pt,
  arc=0pt, outer arc=0pt,
  boxrule=1pt
]
  MLLMs both introduce and flatten subgroup differences.
\end{tcolorbox}

Figure~\ref{fig:pmsv_age} illustrates one form of subgroup mismatch: human PMSV does not differ significantly across age groups, whereas both Gemini and Qwen assign significantly higher PMSV to younger simulated participants. This suggests that MLLMs can introduce subgroup differences that are not observed in the human data. Figure~\ref{fig:pmsv_gender} shows the opposite pattern: female participants have significantly higher PMSV than male participants in the human data, but both MLLMs attenuate this gender difference in their simulations. Thus, MLLMs can also flatten subgroup heterogeneity that is present in the human data. Mixed-effects source-by-group interaction tests confirm that these subgroup patterns differ significantly between human and MLLM-simulated participants ($p<.01$), and the same patterns hold after controlling for sensation-seeking levels.

\begin{table*}[t]
\centering \small
\begin{tabular}{@{}>{\centering\arraybackslash}m{7.7cm}m{9.6cm}@{}}
\hline
\textbf{Video stills} & \textbf{Excerpt from MLLM CoT output} \\ \hline
\includegraphics[width=\linewidth]{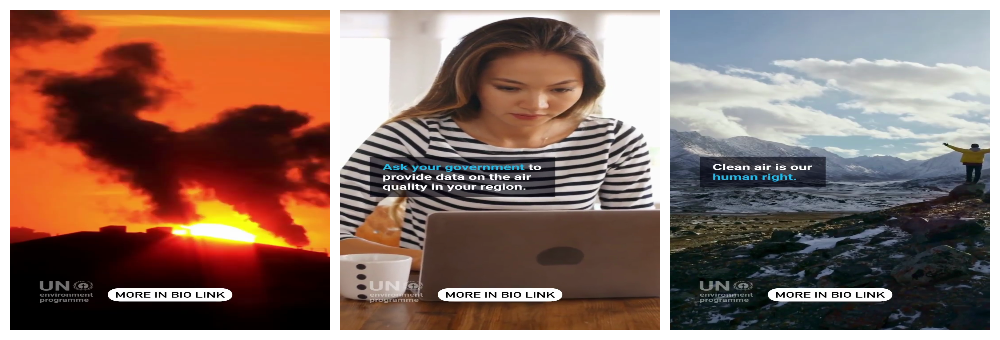}
  & \textit{The video is a public service announcement from the UN Environment Programme regarding the health risks of air pollution. It uses a combination of stock footage (smokestacks, city life, nature) and text overlays to convey statistics and calls to action. The music is somber and steady ... the video follows a very traditional PSA format... the subject of health and 7 million deaths is inherently serious, even if the delivery is subdued.} --- Gemini\\ \midrule
\includegraphics[width=\linewidth]{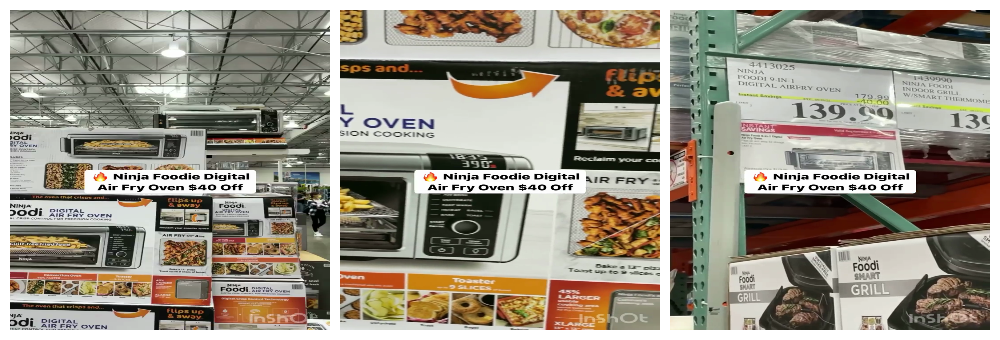}  
  & \textit{The video features a product display in a warehouse setting... with a focus on the Ninja Foodi Digital Air Fry Oven. The video includes a voiceover of the Spider-Man quote, ``With great power comes great responsibility,'' which is a dramatic and emotionally charged line. The background music is intense and cinematic, adding to the emotional and dramatic impact. The visuals are straightforward, showing product packaging and pricing...} --- Qwen \\ \hline
\end{tabular}
\caption{Example CoT output excerpts (see more examples in the Appendix).}
\label{tab:qualitative_study}
\end{table*}

\begin{figure}[t]
    \centering
    \includegraphics[width=0.495\linewidth]{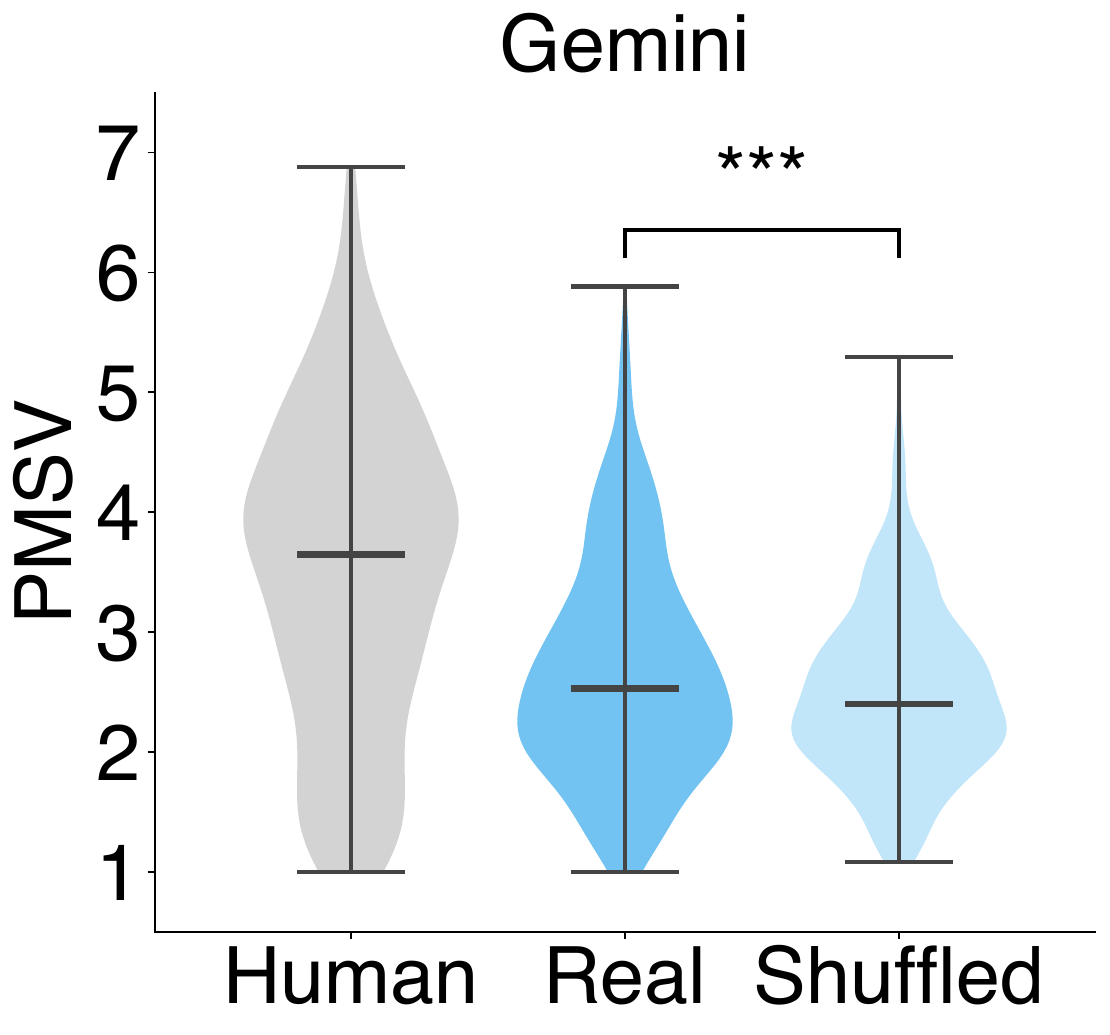}
    \hfill
    \includegraphics[width=0.495\linewidth]{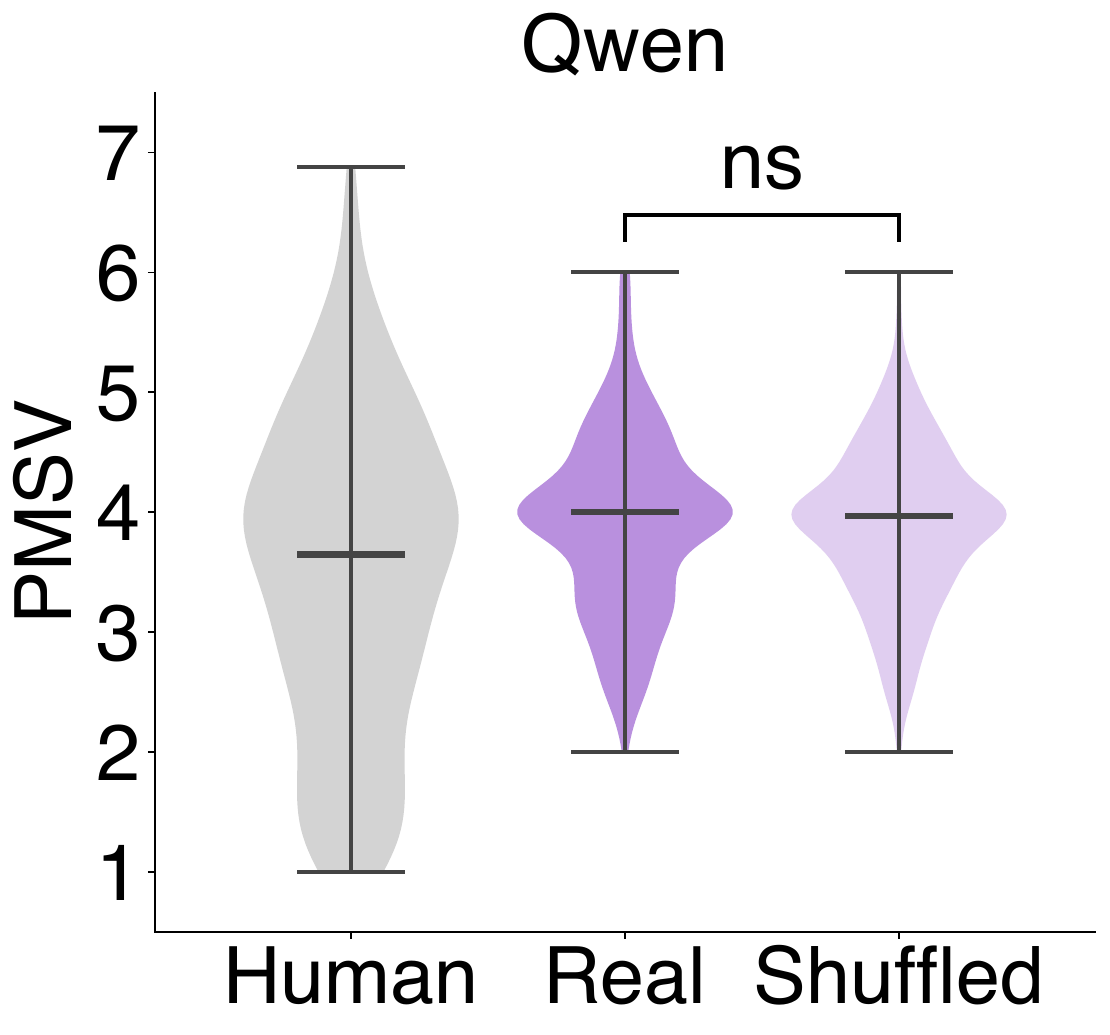}
    \caption{PMSV distributions of MLLM-synthesized participants with real vs shuffled profiles (zero-shot).}
    \label{fig:pmsv_dist_real_vs_shuffle}
\end{figure}

\smallskip
\begin{tcolorbox}[
  title={$\blacktriangleright$ Finding 4},
  detach title,
  coltitle=black,
  fonttitle=\bfseries,
  before upper={%
    \textbf{\tcbtitletext}:\enskip
  },
  top=2pt, bottom=2pt,
  left=2pt, right=2pt,
  arc=0pt, outer arc=0pt,
  boxrule=1pt
]
  MLLMs show limited and inconsistent sensitivity to participant profiles.
\end{tcolorbox}

Here, we examine whether MLLMs' difficulty in simulating subjective sensory engagement may be related to limitations in multimodal video understanding or in profile-conditioned response modeling.

First, qualitative inspection of CoT outputs suggests that both Gemini and Qwen generally produce coherent descriptions of video content by integrating visual, auditory, and textual cues (Table~\ref{tab:qualitative_study}; Appendix). 
Some pre-cutoff videos may overlap with the models' training data, which precludes strong claims about true generalization. However, these qualitative observations, together with prior benchmarks reporting strong multimodal understanding and reasoning in these model families~\cite{xu2025qwen3}, suggest that the observed simulation gap is unlikely to be explained solely by gross failures of video comprehension.

We then tested profile sensitivity by randomly permuting participants' true profiles three times and re-running zero-shot simulations. 
For Gemini, conditioning on shuffled profiles yielded predicted PMSVs of $2.50 \pm 0.72$, compared with $2.67 \pm 0.92$ using real profiles ($p<.001$; $t$-test; Figure~\ref{fig:pmsv_dist_real_vs_shuffle}). Agreement with human participants decreased from 0.14 to 0.02. 
For Qwen, shuffled profiles yielded predicted PMSVs of $3.89 \pm 0.66$, compared with $3.90 \pm 0.75$ using real profiles ($p>.05$; $t$-test; Figure~\ref{fig:pmsv_dist_real_vs_shuffle}). Agreement changed only slightly, from 0.14 to 0.12.
Thus, Gemini shows greater sensitivity to profile perturbation, whereas Qwen's predictions change little when profiles are shuffled.

To address the alternative explanation that the available profiles contain no predictive information, we conducted a supplemental random-forest (RF) analysis. 
Compared with video-only models using video features from \citet{xue2026msv} (Appendix), adding our participant-profile features increased $R^2$ by 0.026 and reduced RMSE by 0.019 when predicting individual human PMSV across 100 repeated participant-level 80/20 train/test splits ($p<.01$; paired $t$-tests). 
Thus, participant profiles contain modest but reliable predictive signal, but current MLLMs show inconsistent sensitivity to this information.

\section{Discussion \& Conclusion}

This work evaluates video-capable MLLMs as synthetic participants in subjective video-based research. We focus on an emerging yet important task: assessing perceived sensory engagement with short videos, operationalized using the communication-theoretic PMSV framework. 
Using PMSV ratings from 673 human participants and profile-conditioned simulations from leading open- and closed-source MLLMs, Gemini 3 Flash and Qwen 3 Omni, we find that current MLLMs remain limited in individual-level fidelity: their agreement with human participants is substantially below a human--human agreement reference.

Their failures are not uniform. Qwen primarily exhibits central-tendency bias, whereas Gemini shows a robust downward mean-shift bias while better preserving the relative ordering of human PMSVs. Prompting has mixed effects: few-shot modestly improves agreement, but CoT does not. Both few-shot and CoT reduce central tendency, yet CoT may worsen mean shift. A possible explanation is that CoT's explicit deliberation and reasoning~\cite{wei2022chain} are misaligned with fast, affective human PMSV judgments and may reinforce initial biases in the absence of external feedback~\cite{huang2023large}.

Although advanced MLLMs perform well on objective tasks, including video understanding and reasoning~\cite{yue2024mmmu,xu2025qwen3}, matching subjective human responses requires more than multimodal comprehension. It also requires calibrated use of participant-level information and sensitivity to heterogeneous human reactions. Our profile-shuffling analysis shows inconsistent sensitivity to participant profiles: Gemini is more affected by profile perturbation, whereas Qwen changes little. A supplemental RF analysis suggests that the profiles contain modest but reliable predictive signal, indicating that weak profile effects are not simply due to entirely uninformative profiles. We interpret this not as evidence that Gemini is superior to Qwen, given their deployment differences, but as evidence that profile utilization is a useful diagnostic dimension and potential direction for improvement.

This study has several limitations. First, we focus on sensory engagement with short Instagram videos, so the results may not generalize to other video genres, platforms, or subjective outcomes. Second, the participant profiles include demographics and sensation seeking but may omit other factors shaping video responses. Third, we evaluate prompting-based methods rather than alternatives such as fine-tuned or task-adapted MLLMs, which may limit simulation fidelity.

Our findings underscore the challenges and complexity of developing MLLMs that can simulate human participants for increasingly common video-based human-subject research. As a first empirical step, this work can be extended in both scale and scope through benchmarks and methods that advance this multimodal, human-centered area.

\section{Broader Impact \& Ethics Statement}

Our study procedures were approved by the university's IRB. Informed consent was obtained from all participants prior to data collection. Participants were recruited via Prolific and compensated at a rate of \$10 per hour. PII was removed from the dataset. The short videos used in this study were from a publicly collected dataset of Instagram Reels~\cite{qian2024convergence,xue2026msv}; no private or restricted content was accessed.
Our findings show that current MLLMs both introduce and flatten subgroup differences such as in age and gender. These limitations pose risks if MLLMs are prematurely adopted as substitutes for human participants, particularly in sensitive domains where biased or flattened responses could lead to misleading conclusions and disproportionately affect underrepresented communities. If adopted without proper validation, MLLM simulations could be misused to bypass genuine human participation, thereby amplifying these biases at scale. We therefore recommend that MLLM simulations be used as complementary tools for refining study protocols and identifying potential risks, rather than replacing genuine human participation.

\bibliography{aaai2026}

\appendix
\section{Prompts} 

Below are our zero-shot, few-shot, and CoT prompts, adapted from \citet{kolluri2025finetuning}.

\begin{tcolorbox}[title=Zero-shot prompt,
  top=2pt, bottom=2pt,
  left=2pt, right=2pt]
\textbf{[SYSTEM]} You are simulating a survey respondent.
Answer exactly as instructed, following the specified
response format without additional commentary.

\smallskip
\textbf{[USER]} \texttt{\color{red}\{Video link\}}

\smallskip
You are a survey respondent with this profile: \\
\texttt{\color{red}\{Demographic and psychographic info.\}}

\smallskip
You just watched a short video. Read the question below and answer exactly as this person would. Follow the response instructions precisely.

\smallskip
Rate this video on each of the following 17 items using a 7-point scale. For each item, assign an integer from 1 to 7 where 1 = the LEFT anchor and 7 = the RIGHT anchor.

\texttt{\color{red}\{PMSV scale items\}}
\end{tcolorbox}

\begin{tcolorbox}[title=Example user message for zero-shot prompting,
  top=2pt, bottom=2pt,
  left=2pt, right=2pt]
\textbf{[USER]} https://generativelanguage.googleapis.com/v1b\\eta/ files/nu11v0d1a4vg

\smallskip
You are a survey respondent with this profile:\\
Age: 29\\
Gender: Female\\
Race/Ethnicity: Asian\\
Education: Associates or technical degree\\
Household Income: \$25,000--\$49,999\\
Sensation-Seeking Score: 3.2 (scale 1--5, higher = more sensation-seeking)

\smallskip
You just watched a short video. Read the question below and answer exactly as this person would. Follow the response instructions precisely. 

\smallskip
Rate this video on each of the following 17 items using a 7-point scale. For each item, assign an integer from 1 to 7 where 1 = the LEFT anchor and 7 = the RIGHT anchor.

\smallskip
[Emotional Arousal]\\
emotional: Unemotional (1) $\leftrightarrow$ (7) Emotional\\
arousing: Not arousing (1) $\leftrightarrow$ (7) Arousing\\
involving: Uninvolving (1) $\leftrightarrow$ (7) Involving\\
exciting: Boring (1) $\leftrightarrow$ (7) Exciting\\
powerful\_impact: Weak impact (1) $\leftrightarrow$ (7) Powerful impact\\
stimulating: Not stimulating (1) $\leftrightarrow$ (7) Stimulating\\
strong\_visual: Weak visuals (1) $\leftrightarrow$ (7) Strong visuals\\
strong\_soundeffect: Weak sound effects (1) $\leftrightarrow$ (7) Strong sound effects

\smallskip
[Dramatic Impact]\\
dramatic: Undramatic (1) $\leftrightarrow$ (7) Dramatic\\
graphic: Not graphic (1) $\leftrightarrow$ (7) Graphic\\
creative: Not creative (1) $\leftrightarrow$ (7) Creative\\
goosebump: Didn't give me goosebumps (1) $\leftrightarrow$ (7) Gave me goosebumps\\
intense: Not intense (1) $\leftrightarrow$ (7) Intense\\
strong\_soundtrack: Weak soundtrack (1) $\leftrightarrow$ (7) Strong soundtrack

\smallskip
[Novelty]\\
novel: Ordinary (1) $\leftrightarrow$ (7) Novel\\
unique: Common (1) $\leftrightarrow$ (7) Unique\\
unusual: Usual (1) $\leftrightarrow$ (7) Unusual
\end{tcolorbox}

\begin{tcolorbox}[title=Modified system message for few-shot prompting,
  top=2pt, bottom=2pt,
  left=2pt, right=2pt] 
\textbf{[SYSTEM]} You are simulating a survey respondent. Answer
exactly as instructed, following the specified response format
without additional commentary. Use the examples to calibrate the scale; note how the ratings relate to what you see in each video.

\smallskip
=== Example 1 ===\\
\texttt{\color{red}\{Video 1 link\}}\\
Respondent 1 profile: \\
\texttt{\color{red}\{Demographic and psychographic info.\}}\\
Survey ratings for the video above:\\
\texttt{\color{red}\{PMSV scale items and respondent ratings\}}

\smallskip
=== Example 2 ===\\
\texttt{\color{red}\{Video 2 link\}}\\
Respondent 2 profile: \\
\texttt{\color{red}\{Demographic and psychographic info.\}}\\
Survey ratings for the video above:\\
\texttt{\color{red}\{PMSV scale items and respondent ratings\}}

\smallskip
=== Example 3 ===\\
\texttt{\color{red}\{Video 3 link\}}\\
Respondent 3 profile: \\
\texttt{\color{red}\{Demographic and psychographic info.\}}\\
Survey ratings for the video above:\\
\texttt{\color{red}\{PMSV scale items and respondent ratings\}}

\end{tcolorbox}

\begin{tcolorbox}[title=Modified user message for CoT prompting,
  top=2pt, bottom=2pt,
  left=2pt, right=2pt]
\textbf{[USER]} \texttt{\color{red}\{Video link\}} 

\smallskip
You are a survey respondent with this profile:\\
\texttt{\color{red}\{Demographic and psychographic info.\}}

\smallskip
You just watched a short video. Read the question below and answer exactly as this person would. Follow the response instructions precisely.

\smallskip
Rate this video on each of the following 17 items using a 7-point scale. For each item, assign an integer from 1 to 7 where 1 = the LEFT anchor and 7 = the RIGHT anchor.

\texttt{\color{red}\{PMSV scale items\}}

\smallskip
Before giving the final ratings, follow these steps:

Step 1: Summarize the video's main characteristics.\\
Step 2: Explain how this respondent would likely react to it.\\
Step 3: Explain how you map that reaction to the rating scales.\\
Step 4: Give the final ratings.

\smallskip
Use this exact output format:

Reasoning:\\
1.\ \ldots\\
2.\ \ldots\\
3.\ \ldots

Final ratings:
\end{tcolorbox}

\section{Additional Experimental Setup Details}

\begin{table*}[t]
\centering
\begin{tabular}{rl}
 & \textbf{Items} \\ \midrule
\textbf{Strongly disagree (1)} & \textit{I would like to explore strange places.} \\
\textbf{to strongly agree (5)} & \textit{I would like to take off on a trip with no pre-planned routes or timetables.} \\
& \textit{I get restless when I spend too much time at home.}  \\
& \textit{I prefer friends who are excitingly unpredictable.}  \\
& \textit{I like to do frightening things.} \\
& \textit{I would like to try bungee jumping.} \\
& \textit{I like wild parties.} \\
& \textit{I would love to have new and exciting experiences, even if they are illegal.} \\
\end{tabular}
\caption{The brief sensation-seeking scale~\cite{hoyle2002reliability}.}
\label{table:ss_measure}
\end{table*}

\paragraph{Sensation seeking} We measured sensation seeking using the brief sensation-seeking scale~\cite{hoyle2002reliability}. 
Participants indicated the extent to which they agreed or disagreed with eight statements on a 5-point scale (Table~\ref{table:ss_measure}). We computed the sensation-seeking score as the mean of the eight items; therefore, the resulting value remains within the original 1 to 5 range~\cite{hoyle2002reliability}. In our dataset, sensation-seeking scores ranged from 1.00 to 4.88.

\begin{table}[t]
\centering 
\begin{tabular}{rlr} 
 & \textbf{Items (Count)}  \\ \midrule
\textbf{Gender} & Male (318) \\
\textbf{} & Female (337) \\
\textbf{} & Other (15) \\
\textbf{} & Prefer not to say (3) \\ 
\textbf{Ethnicity} & White (466) \\ 
\textbf{} & Black / African American (88) \\
\textbf{} & Asian (43) \\
\textbf{} & Other (32) \\
\textbf{} & Multiracial (38) \\
\textbf{} & Prefer not to say (6) \\ 
\textbf{Education} & Some high school or less (5) \\
\textbf{} & High school diploma or GED (77) \\
\textbf{} & Some college, but no degree (161) \\
\textbf{} & Associates or technical degree (82) \\
\textbf{} & Bachelor's degree (244) \\
\textbf{} & Graduate or professional degree (102) \\
\textbf{} & Prefer not to say (2) \\ 
\textbf{Household} & Less than \$25,000 (98) \\
\textbf{income} & \$25,000 – \$49,999 (155) \\
\textbf{} & \$50,000 – \$74,999 (150) \\
\textbf{} & \$75,000 – \$99,999 (91) \\
\textbf{} & \$100,000 – \$149,999 (100) \\
\textbf{} & \$150,000 or more (56) \\
\textbf{} & Prefer not to say (23) \\ 
\end{tabular}
\caption{Participant demographics by survey item. Age was recorded as integer values and ranged from 19 to 83 years.}
\label{tab:sample_characteristics}
\end{table}

\paragraph{Participant demographics}  Table~\ref{tab:sample_characteristics} presents the distribution of participant demographics by survey item.

\begin{table}[t]
\centering
\begin{tabular}{rl} 
 & \textbf{Items (1--7)} \\ \midrule
\textbf{Emotional} & Emotional--unemotional \\
\textbf{arousal} & Arousing--not arousing \\
 & Involving--uninvolving \\
 & Exciting--boring \\
 & Powerful--weak impact \\
 & Stimulating--not stimulating \\
 & Strong--weak visuals \\
 & Strong--weak sound effects \\
\textbf{Dramatic} & Dramatic--undramatic \\
\textbf{impact} & Graphic--not graphic \\
 & Creative--not creative \\
 & Didn't give--gave me goosebumps \\
 & Intense--not intense \\
 & Strong--weak soundtrack \\
\textbf{Novelty} & Novel--ordinary \\
 & Unique--common \\
 & Unusual--usual \\ 
\end{tabular}
\caption{PMSV scale~\cite{palmgreen2002perceived}}
\label{table:pmsv_measure}
\end{table}

\paragraph{PMSV} 
Table~\ref{table:pmsv_measure} presents the PMSV scale developed by \citet{palmgreen2002perceived}. In our dataset, Cronbach's alpha was 0.95 for humans, 0.96 for Gemini, and 0.97 for Qwen (zero-shot).

\paragraph{Implementation details}
All experiments were run on an Intel Xeon W5-2565X CPU with 64 GB RAM and an NVIDIA RTX 6000 Ada GPU with 48 GB VRAM. We used Gemini 3 Flash via its API with default hyperparameters, and we deployed Qwen 3 Omni locally using NF4 precision. For Qwen 3 Omni, input videos were processed at 16 FPS with a resolution of 168x112, and the few-shot example videos were processed at 8 FPS.

\begin{table}[t]
\centering 
\begin{tabular}{rll}
 & \textbf{Gemini} & \textbf{Qwen} \\ \midrule
\textbf{Zero-shot} & 1.27 & 1.12 \\
\textbf{Few-shot} & 1.17 & 1.12 \\
\textbf{CoT} & 1.41 & 1.31 \\
\end{tabular}
\caption{MAE in PMSV between human participants and MLLM-synthesized participants across prompting methods.}
\label{tab:mae_pmsv} 
\end{table}

\begin{table*}[t]
\centering
\begin{tabular}{@{}>{\centering\arraybackslash}m{6cm}m{11.2cm}@{}}
\textbf{Video stills} & \textbf{Excerpt from MLLM CoT output} \\ \midrule
\includegraphics[width=\linewidth]{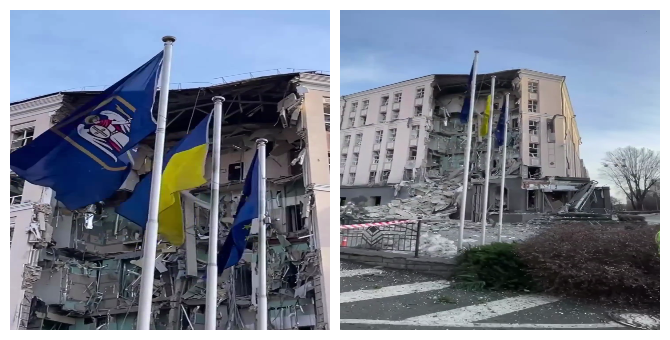}
 & {\textit{The video depicts the aftermath of a severe bombing or missile strike on a large building, with Ukrainian and EU flags visible. The scene is one of significant destruction and rubble, accompanied by the sounds of sirens and ambient noise...}} \\ 
 \includegraphics[width=\linewidth]{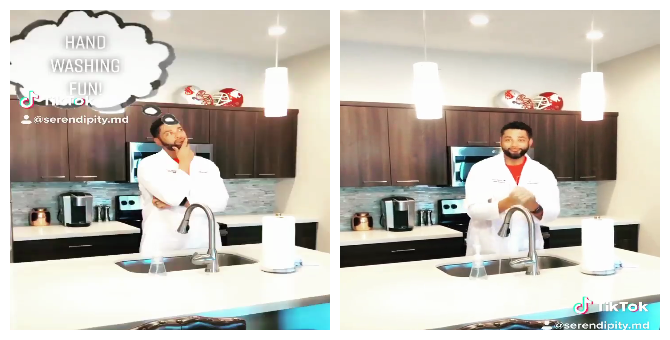}
 & {\textit{The video features a man in a white lab coat, likely a doctor, in a kitchen setting, performing a handwashing demonstration. The video is set to a hip-hop/rap song with a strong beat, creating a contrast between the serious medical content and the energetic music. The man is dancing while washing his hands, which adds a humorous and unconventional element to the video. The video is likely intended to be educational but with a fun, engaging twist...}} \\ 
\end{tabular}
\caption{Additional example CoT output excerpts.}
\label{tab:qualitative_study_2}
\end{table*}

\section{Additional Experimental Results}

\paragraph{Human--human agreement} 
For each video, we randomly sampled two human participants with matching demographic and psychographic features, then randomly assigned their PMSVs to two groups. We then computed the PMSV agreement between the two groups. We repeated this procedure 10{,}000 times and report the mean agreement of 0.29, with a 95\% CI of [0.269, 0.305].

\paragraph{MAE} Table~\ref{tab:mae_pmsv} reports the MAE in PMSV between human participants and MLLM-synthesized participants across prompting strategies.

\paragraph{LMMs} We used LMMs to compare human and MLLM ratings. For the mean-shift analysis, rating source was entered as a fixed effect, and participant was entered as a random intercept to account for repeated ratings from the same participant. The corresponding fixed-effect test assessed whether mean ratings differed between MLLM and human ratings. To assess central tendency bias, we computed each rating's squared deviation from the grand mean of ratings within its source and fit the same LMM structure to these squared deviations. The corresponding fixed-effect test assessed whether rating variability differed between MLLM and human ratings. All models were fit using restricted maximum likelihood (REML).

\paragraph{Qualitative analysis}
Table~\ref{tab:qualitative_study_2} presents additional examples in which Gemini and Qwen generally describe video content coherently by integrating multimodal cues, including visual, auditory, and textual information.

\paragraph{Video features} We adapted the 20 video features from \citet{xue2026msv}, which were shown to be effective for predicting a video's sensation value. 
These video features include three audio features (loudness, tempo, sound brightness) and seventeen visual features. The visual features consist of four static features (brightness, color warmness, color saturation,  visual complexity); two dynamic features (shot count/s, shot duration/s); nine static face features (size, age, gender, six discrete emotions); and two dynamic face features (face count/s and face duration/s).

\end{document}